\documentclass{article}

\usepackage{graphicx}
\usepackage{amsmath}
\usepackage{amsfonts}
\usepackage{bm}

\newenvironment{myfigure}
{\begin{figure}\begin{center}\begin{minipage}{0.85\linewidth}\begin{center}}
{\end{center}\end{minipage}\end{center}\end{figure}}

\newcommand{\mycaption}[1]{\caption{{\protect\footnotesize #1}}}

\begin{document}



\title{Flow of thin liquid films on chemically structured substrates}

\author{\textsc{Markus Rauscher} and \textsc{S. Dietrich}\\[0.5ex]
\textit{Max-Planck-Institut f{\"u}r Metallforschung}, \\
\textit{Heisenbergstr. 3, 70569 Stuttgart, Germany}, and\\
\textit{Institut f{\"u}r Theoretische und Angewandte Physik,}\\ 
\textit{Universit{\"a}t Stuttgart, Pfaffenwaldring 57, 70569 Stuttgart,
Germany}}
\date{(\today)}
\maketitle

\abstract{
Chemically patterned surfaces are of significant interest in the context of
microfluidic applications. Miniaturization of such devices will eventually 
lead to structures on the nano-scale. Whereas on the micron scale purely
macroscopic descriptions of liquid flow are valid, on the nanometer scale 
long-ranged inter-molecular interactions, thermal fluctuations such as
capillary waves, and finally the molecular structure of the liquid become
important.
We discuss the most important conceptual differences between
flow on chemically patterned substrates on the micron scale and on the 
nanometer scale. These concern the structure of the triple line, the type
of interactions between neighboring liquid flows, and the influence of the
molecular structure of the liquid on the flow. We formulate four design
issues for nanofluidics related to channel width, channel separation, and 
channel bending radius, and conclude with a discussion
of the relevance of the conceptual differences between the micron scale and
the nanometer scale for these issues.
}

\section{Introduction}

In recent years substantial efforts have been invested in
miniaturizing chemical processes by building microfluidic systems. The
``lab on a chip concept'' integrates a great variety of chemical and physical
processes into a single device \cite{giordano01,mitchell01,stone01} 
in a similar way as an integrated circuit
incorporates many electronic devices into a single chip.
 These microfluidic
devices do not only allow for cheap mass production but they 
can operate with much smaller quantities of reactants and
reaction products than standard laboratory equipments. This is particularly
important for rare and expensive substances such as certain biological
substances and for toxic or explosive materials. 
Even though most available microfluidic devices today have micron sized
channels further miniaturization is leading towards the nano-scale.
Besides meeting technical challenges, new theoretical concepts are needed 
to understand the basic physical processes underlying this new technology.
Whereas the ultimate limits for the miniaturization of electronic devices
are set by quantum fluctuations, in a chemical chip these limits are
determined by thermal fluctuations and can be explored by methods of
classical statistical mechanics.

There are two main lines of development for microfluidic systems. The
first one encompasses systems with closed channels. A common technique to
produce these devices is to cast poly(dimethylsiloxane) (PDMS) over a
topographically structured master, to peel off the polymer after curing, and
to seal the resulting topographically structured material onto a flat surface.
This way rather complicated devices including valves and pumps have been
fabricated \cite{jeon02}\/. However, closed channel systems have the
disadvantage that they can be easily clogged by solute particles such as
colloids or large bio-polymers. 

The second type of systems are open with a free liquid-vapor interface and
the fluid is not confined by physical but by chemical walls.  The idea is
that the liquid will be guided by lyophilic stripes on an otherwise 
lyophobic substrate. There are two sub-types of this technique: using a
single chemically patterned substrate \cite{darhuber01,gau99} (see also
Fig.~\ref{samplefig}) or a chemically patterned slit pore
\cite{handique00,lam02,zhao02}, respectively. The substrate surfaces can be
structured chemically by printing or photographical techniques.
All the techniques are confined to two dimensions.

\begin{myfigure}
\includegraphics[width=\linewidth]{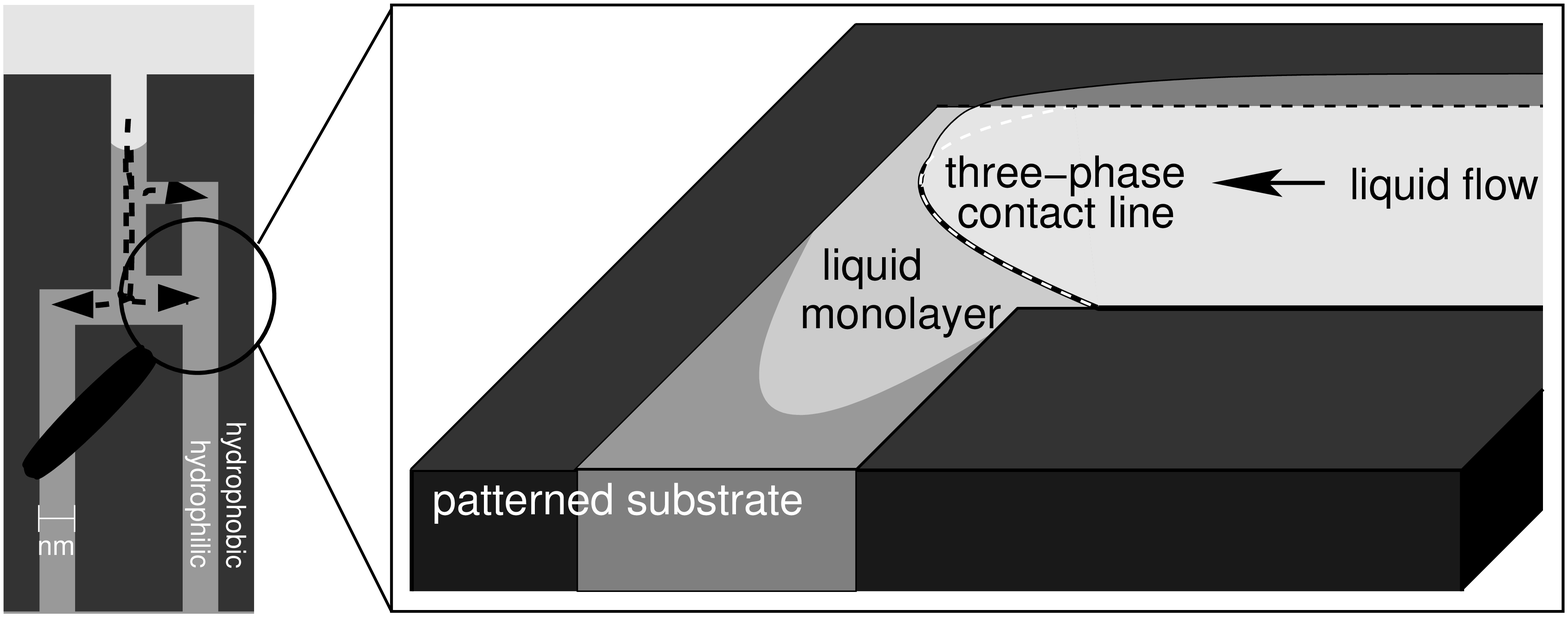}
\mycaption{\label{samplefig} Schematic picture of liquid flow guided by 
so-called chemical walls on a
chemically patterned substrate. The liquid moves along lyophilic stripes
(light grey) on a lyophobic substrate (dark grey)\/. The liquid layer ends
in a three-phase contact line and a liquid monolayer (the precursor film) 
spreads ahead of it.}
\end{myfigure}

Liquids in contact with chemically structured substrates are not
only of technological use but also enjoy high interest in basic research. 
Wetting on chemically structured substrates
has been studied extensively on all length scales. For example
adsorption on stripe patterns shows a rich morphology on the macro-scale
\cite{lenz00,brinkmann02} as well as on the nano-scale
\cite{dietrich99,bauer00}\/. In this context
flow of liquids has also been studied to a large extend. Most of this
research has been focused on homogeneous substrates, in particular on dynamic
contact angles, dewetting of thin films, and flow of thin films in many
circumstances. For a review see Ref.~\cite{oron97}\/.

In the following we first describe briefly those equilibrium features of 
wetting on chemically patterned substrates which are most relevant for 
microfluidic applications. After discussing the characteristics of thin
film flow on homogeneous substrates, we state the three main conceptual
differences between micro-scale and nano-scale fluid transport in chemical
channels. We conclude with formulating four design principles for
nanofluidic devices.

\section{Wetting on chemically patterned substrates}

Equilibrium wetting phenomena on chemically patterned substrates have been 
analyzed theoretically in great detail on both the macroscopic and the
microscopic scale. 
Microscopic theories, such as the successful density functional
theory, take into account the finite range of inter-molecular attractions
and short-ranged repulsions explicitly \cite{evans90}\/. Density functional
theories do not only allow to study the order of wetting transitions and
the equilibrium shape of the wetting film but also the
detailed microscopic structure of the liquid in the vicinity of the
substrate and at the liquid-vapor interface \cite{dietrich98}\/.

In macroscopic theories, however, the inter-molecular interactions are
approximated by local descriptions. For a wetting film this means that the
free energy of the film is described by a bulk term proportional to the
volume of the fluid, a surface tension term proportional to the area of the
liquid-vapor interface, an interface term proportional to the area of the
liquid-substrate interface, and a line tension term proportional to the
length of the three phase  contact line between liquid, vapor, and
substrate.  In most studies the line tension term is neglected.
Macroscopic theories have been used to describe the shape of
droplets on homogeneous and structured substrates
\cite{lenz00,brinkmann02}\/. The equilibrium droplet shape in chemically
patterned slit-like pores has also been studied with the same technique 
\cite{valencia01}\/.

It is a great challenge to describe the intermediate scale between the 
microscopic 
and the macroscopic one. In most cases it is impossible to obtain
analytical results from microscopic theories 
and numerical calculations are prohibitive for large systems.
Current research focuses on extending the scope of macroscopic theories
down to the meso-scale by incorporating microscopic effects such as the
wave-length dependence of the surface tension \cite{mecke99} and detailed
properties of the three phase contact line \cite{bauer99b}\/.

\section{Flow of thin liquid films}

Flow of thin liquid films on homogeneous substrates has been studied
extensively, in particular the motion of the three-phase contact line,
dewetting of thin films, the stability of falling liquid films, and
Marangoni flow. 
Apart from a few molecular dynamics simulations, in most studies the liquid
flow is described in terms of meso-scale hydrodynamics. This means that
the hydrodynamic equations are augmented with long-ranged liquid-substrate
interactions and hydrodynamic slip of the liquid at the substrate (with
slip lengths on the nanometer scale). For a review see Ref.~\cite{oron97}\/. In
a phase-field description even compressibility effects and the finite width
of the liquid-vapor interface have been taken into account
\cite{pismen02}\/. The finite interface width in the phase field models
is the only signature of thermal fluctuations taken into account in
hydrodynamic thin film models up to now.

The main analytic tool in this approach is the so-called lubrication
approximation, which is a small-gradient expansion for the film thickness
and which leads to a fourth order in space, first order in time parabolic
partial differential equation for the film thickness. This equation is also
often referred to as the thin film equation and it has been successful in
describing dewetting processes quantitatively \cite{becker02}\/.  
Flow over chemical substrate
inhomogeneities has been studied in the context of dewetting of unstable
films \cite{kargupta01a} but to little extent for the actual situations 
relevant to microfluidics and nanofluidics \cite{brusch02,kondic02b}\/. 

\section{Conceptual differences in microfluidics and nanofluidics}

In this section we state three important conceptual differences
between microfluidics and nanofluidics. 

\subsection{Triple line and chemical steps}
 
\begin{myfigure}
\includegraphics[width=\linewidth]{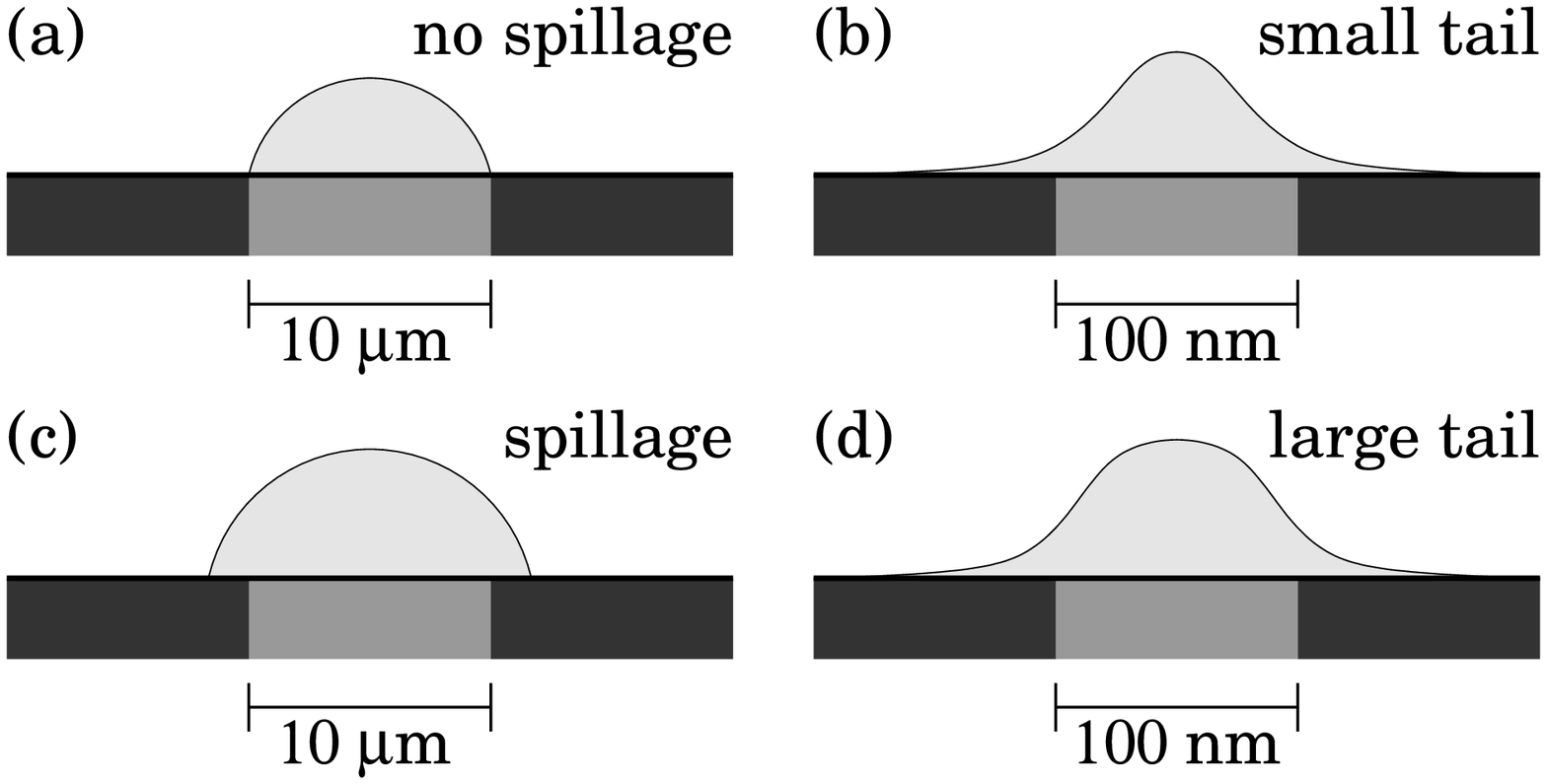}
\mycaption{\label{spillagefig} Spill-over of a liquid channel on a
micron-sized chemical strip
(left column) as compared to a nano-scale strip (right column)\/. 
Dark grey marks the lyophobic areas and light grey the lyophilic areas.
At the micron scale one can clearly distinguish between 
no spillage (a) and spillage (c), whereas for nano-channels one can only
distinguish between small tails (b) and large tails (d) of the lateral
liquid distribution\/.}
\end{myfigure}

One of the most critical issues in open microfluidic systems is to keep the
liquid in the desired areas such as channels, reactors, and reservoirs. On a
macroscopic scale the liquid will stay on the lyophilic channels
for low filling and the three phase contact line will lie on the channel
area or it will be pinned at the chemical step. Spill-over onto the
lyophobic areas occurs once the contact angle of the liquid at the chemical
step exceeds the advancing contact angle $\alpha_a$ on the lyophobic area
(which is in general larger than the equilibrium contact angle $\alpha_e$
due to surface defects)\/. 

On the nano-scale the situation is quite different. First, the concepts
of a contact line and contact angle have to be revised. 
A sharp contact line is replaced by a smooth transition from a
mesoscopic wetting film to the precursor film
\cite{bauer00,koch95,getta98,bauer99c,bauer99a} which is only some few
molecular diameters thick and spreads ahead of the main portion of the
moving liquid. Moreover even an atomically sharp boundary
between lyophilic and lyophobic areas on the substrate will lead to a
smooth lateral variation of the interaction potential between the liquid
particles and the substrate.

Thus the macroscopic and sharp criterion for a liquid staying on a chemical
channel, namely whether the triple line crosses the channel boundary or
not, becomes fuzzy at the nano-scale. Since there is always
a certain amount of liquid on the lyophobic part of the substrate, one
has to address the issue which fraction of the liquid is outside the 
channels rather than whether there is liquid outside the channels.

\subsection{Interaction between neighboring channels}

\begin{myfigure}
\includegraphics[width=\linewidth]{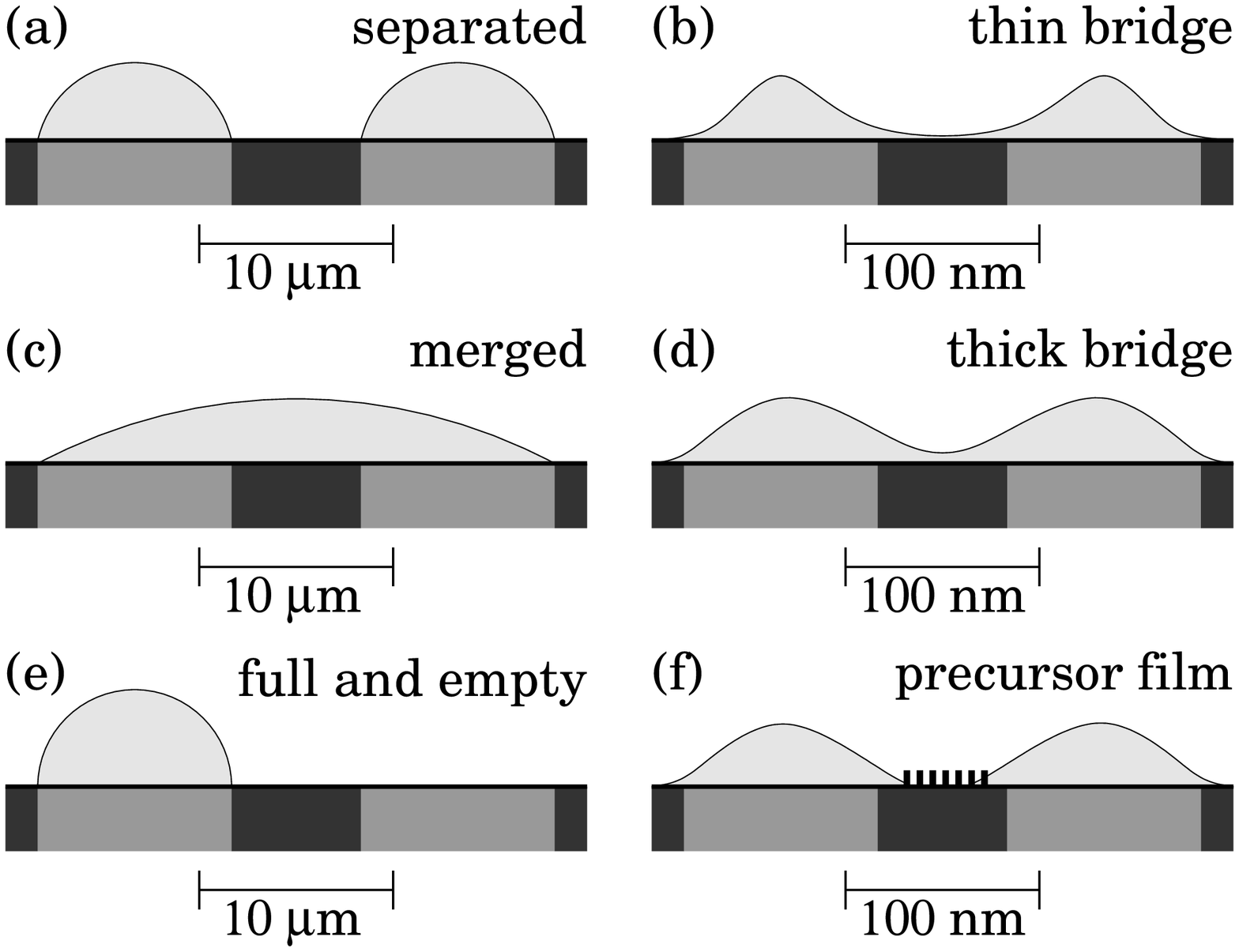}
\mycaption{\label{interactionfig} Interaction between neighboring channels at
the micron scale (left column) and at the nano-scale (right column)\/. 
Dark grey marks the lyophobic areas and
light grey the lyophilic areas. At the micron scale the
flows are either well separated (a) or merged (c). Even a filled channel
next to an empty one (e) is possible. At the nano-scale the tails of the
liquid in
the channels will merge and thus form a thin (b) or thick (d) bridge
through which material can be interchanged. The thickness of the bridge 
is that of a monolayer in the case of a precursor film (f)\/. }
\end{myfigure}

The question of spillage is of course closely related to the question of
interaction between neighboring channels. Macroscopically (neglecting
evaporation and recondensation) two neighboring channels will
interact once the two liquid films merge. One has to keep in mind
that in a macroscopic description an empty and a filled channel next to
each other (Fig.~\ref{interactionfig}(e)) or two filled but not interacting
channels (Fig.~\ref{interactionfig}(a)) can be metastable
states. In the first case two equally filled channels and in the second
case a liquid bridge can be the equilibrium state.
How such configurations are affected by flow has not been investigated yet.

On the nano-scale, however, tails of the liquid from two neighboring
channels can leak onto the lyophobic area between the channels.
If the channels are too close these tails will 
overlap and thus form a bridge (see
Fig.~\ref{interactionfig}(b) and (d))\/. Liquid can flow through such a
bridge and particles immersed in the fluid can diffuse through these
bridges. Thus keeping the two flows separated is
a question of time-scales and not a question that can be answered
definitively.

Also the presence of a monomolecular precursor film on the lyophobic 
area can lead to an exchange of molecules (see
Fig.~\ref{interactionfig}(f))\/.
Macroscopically this is certainly negligible but since the ratio between the
film thickness in the channel and the thickness of the precursor film can be
of the order of $10$, this can be significant at the nano-scale. 

Fluctuations also become more important at small scales. In the absence of
a precursor film on the area between the channels, a filled channel next to
an empty one can be a metastable state as in the macroscopic case 
discussed above. But the energy barrier which has to be overcome to
connect the two channels is much smaller than for macroscopic distances and
much smaller fluctuations of the film height are needed.

Not only the liquid-substrate interactions but also the liquid-liquid
interaction has a range of up to $100$~nm\/. Thus the direct interaction
between liquid streams in parallel channels will influence the flow and
also the thermal fluctuations.


\subsection{Flow and the atomistic structure of liquids}

\begin{myfigure}
\includegraphics[width=\linewidth]{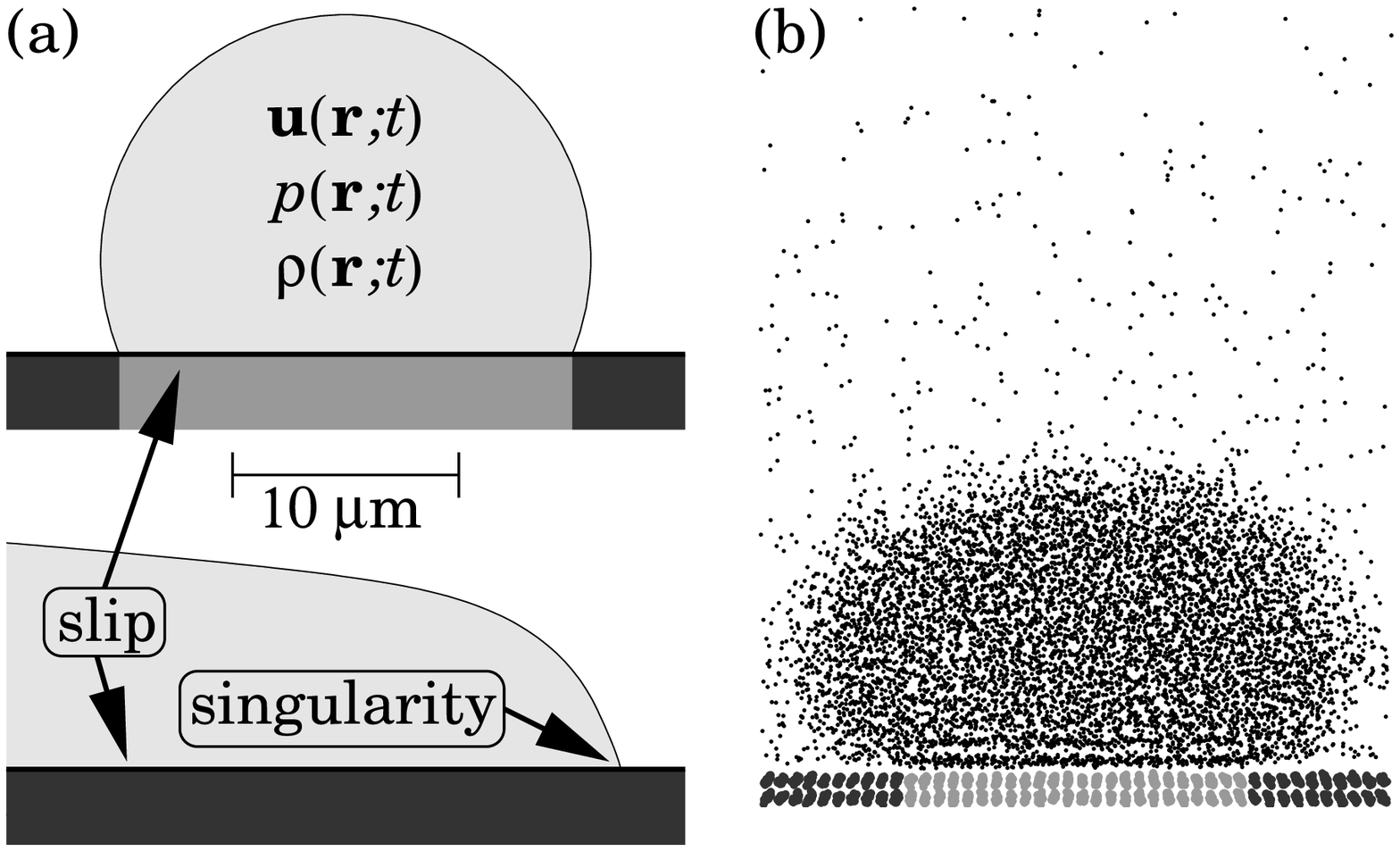}
\mycaption{\label{atomisticfig} On the micron scale (a) the flow can be
described hydrodynamically, \textit{i.e.}, the liquid is described by the
local flow field $\mathbf{u}(\mathbf{r};t)$, the local pressure
$p(\mathbf{r};t)$, and the density $\rho(\mathbf{r};t)$\/. On the nano-scale
(b) the atomic structure of the fluid and the substrate cannot be ignored
\protect\cite{koplikX}\/.}
\end{myfigure}

The conclusions in the last two subsections are mostly based on
quasi-static considerations. Transport mechanisms and dynamic properties 
have not been discussed. Experience tells that down to length scales of
about $1$--$10$~nm hydrodynamic theories provide a quite
good description of liquid flow. However, even at the micron scale
atomistic properties of the liquid show up via the slip length at the
liquid-substrate interface and via the details of the regularization of the
stress singularity at the moving triple line (see
Fig.~\ref{atomisticfig}(a))\/. 

In nanofluidic systems there is a window of length scales within which
long-ranged intermolecular forces play a role and hydrodynamics is
still applicable. This window is centered around film thicknesses of about 
$100$~nm\/.
Below this length scale the atomistic structure of the liquid comes into
play. Currently only molecular dynamics simulations explore this atomic 
length scale region (see for example Fig.~\ref{atomisticfig}(b))\/.

\section{Design issues for nanofluidics}

For liquid flow inside chemically patterned micron-sized channels several
design issues have been addressed \cite{zhao02}\/. Due to the 
conceptual differences between flow on the micron scale and flow on the
nano-scale the answers given for the micron scale cannot
be transferred directly to the nano-scale. However, the basic issues are
the same: (1) How much liquid can a chemical channel contain
before considerable spillage onto the lyophobic areas occurs? (2) How wide
must the channels be to support flow? (3) How small can the radius of
curvature of bends in the channel be? (4) What is the minimum distance
between liquid streams below which they interact?

In micron scale channels these issues can be addressed by considering
only surface tensions, contact angles, and line tensions. On the
nano-scale the situation is more complex. In particular the details of the
interaction between liquid and substrate, the interaction among the liquid
molecules, the influence of the atomistic structure of the fluid on the
transport properties, and thermal fluctuations have to be taken into
account in addition.

\end{document}